\documentclass[10.5pt,superscriptaddress,longbibliography,twocolumn]{revtex4-2}
\usepackage{amsmath,amssymb,amsfonts,float,graphics,color,verbatim,tabularx,bm,multirow,appendix,wasysym}
\usepackage[utf8]{inputenc}
\usepackage[T1]{fontenc}
\usepackage{xcolor}
\usepackage{dsfont}

\usepackage{textcomp}
\usepackage{times}
\usepackage{graphicx}
\usepackage{float}
\usepackage{latexsym,euscript}
\usepackage{color}
\usepackage{subfigure}
\usepackage[bottom]{footmisc}

\usepackage[colorlinks=true,linkcolor=blue,citecolor=blue,urlcolor=blue]{hyperref}
\usepackage{soul}
\usepackage[normalem]{ulem}
\usepackage{mathrsfs}
\usepackage{xfrac}
\usepackage{xspace}
\usepackage{textcomp}
\usepackage{yfonts}
\usepackage{verbatim}
\usepackage{braket}
\usepackage{tikz}
	\usetikzlibrary{calc}
	\usetikzlibrary{shapes.multipart}

\usepackage{url}
\newcommand{\comments}[1]{}

\def\Z{\mathbb{Z}}

\newcommand{\stkout}[1]{\ifmmode\text{\sout{\ensuremath{#1}}}\else\sout{#1}\fi}

\begin{document}

\title{Measuring R\'enyi entanglement entropy with high efficiency and precision in quantum Monte Carlo simulations}

\author{Jiarui Zhao}
\affiliation{Department of Physics and HKU-UCAS Joint Institute of Theoretical and Computational Physics, The University of Hong Kong, Pokfulam Road, Hong Kong SAR, China}

\author{Bin-Bin Chen}
\affiliation{Department of Physics and HKU-UCAS Joint Institute of
	Theoretical and Computational Physics, The University of Hong Kong,
	Pokfulam Road, Hong Kong SAR, China}

\author{Yan-Cheng Wang}
\affiliation{Beihang Hangzhou Innovation Institute Yuhang, Hangzhou 310023, China}

\author{Zheng Yan}
\affiliation{Department of Physics and HKU-UCAS Joint Institute of Theoretical and Computational Physics, The University of Hong Kong, Pokfulam Road, Hong Kong SAR, China}
\affiliation{State Key Laboratory of Surface Physics and Department of Physics, Fudan University, Shanghai 200438, China}

\author{Meng Cheng}
\email{m.cheng@yale.edu}
\affiliation{Department of Physics, Yale University, New Haven, CT 06520-8120, U.S.A}

\author{Zi Yang Meng}
\email{zymeng@hku.hk}
\affiliation{Department of Physics and HKU-UCAS Joint Institute of Theoretical and Computational Physics, The University of Hong Kong, Pokfulam Road, Hong Kong SAR, China}

\begin{abstract}
	We develop a nonequilibrium increment method in quantum Monte Carlo simulations to obtain the R\'enyi entanglement entropy of various quantum many-body systems with high efficiency and precision. To demonstrate its power, we show the results on a few important yet difficult $(2+1)d$ quantum lattice models, ranging from the Heisenberg quantum antiferromagnet with spontaneous symmetry breaking, the quantum critical point with O(3) conformal field theory (CFT) to the toric code $\Z_2$ topological ordered state and the Kagome $\Z_2$ quantum spin liquid model with frustration and multi-spin interactions. In all these cases, our method either reveals the precise CFT data from the logarithmic correction or extracts the quantum dimension in topological order, from the dominant area law in finite-size scaling, with very large system sizes, controlled errorbars and minimal computational costs. Our method therefore establishes a controlled and practical computation paradigm to obtain the difficult yet important universal properties in highly entangled quantum matter.
\end{abstract}

\date{\today}
\maketitle

Entanglement entropy (EE) is a physical quantity that measures the quantum entanglement in an interacting many-body system. The scaling form of the entanglement entropy  contains universal information of the system and can be used to characterize quantum phases and phase transitions in interacting lattice models. For these reasons, the understanding and computation of EE attract much attention from the condensed matter, quantum information and high-energy communities.  Significant theoretical progress have been made to determine the universal terms in the EE in different kinds of quantum phases, including quantum critical systems, gapped phases in (2+1)d, and systems with spontaneous continuous symmetry breaking~\cite{CARDY1988,Calabrese_2004,Fradkin2006,Casini2006,Kitaev2006,Levin2006,Wolf2006,YCLin2007,RongYu2008,Hastings2010,Metlitski2011,Isakov2011,Jiang2012,Casini2012,Swingle2012,Kovacs2012,Inglis2013,InglisNJP2013,KallinPRL2013,Luitz2014,KallinJS2014,Helmes2014,Laflorencie2016}, and on the computational front the Quantum Monte Carlo methods have proven to be a reliable way to numerically measure the R\'enyi entanglement entropy\cite{DEmidio2020,Hastings2010,Helmes2014,Grover2013,Humeniuk2012,JRZhao2021,Luitz2014,Out-of-equilibrium-2017}.

We now briefly review what has been known about the scaling form of EE in (2+1)d, which is the main focus of this work. Suppose the system is partitioned into regions $A$ and $\bar{A}$, and denote the length of the boundary of $A$ by $l$. For critical or gapped systems, the entanglement entropy (both von Neumann and R\'enyi) takes the following form when $A$ is sufficiently large:
\begin{equation}
	S=a l - s \ln(l) - \gamma + O(l^{-1}).
\label{eq:eq1}
\end{equation}
Here $a$ is a non-universal area law coefficient. The coefficient of the logarithmic correction $s$ is a universal number determined by the  geometric properties of the partition (i.e. if $A$ is a rectangle surrounded by $\bar{A}$) and intrinsic properties of the underlying physical system (i.e. central charge of the conformal field theory at certain critical points)~\cite{Fradkin2006, Casini2006}. When $s=0$, the constant $\gamma$ becomes universal and when the system is gapped, it becomes independent of the shape of $A$ and is known as the topological entanglement entropy (TEE). In that case the value of $\gamma$  is given by the logarithm of the total quantum dimension of the underlying topological order~\cite{Kitaev2006,Levin2006}.

It is of great interest to obtain the values of the universal coefficients $s$ and $\gamma$ since they provide access to universal properties of the system that are hard to measure otherwise. Extracting $s$ or $\gamma$ requires finite-size scaling with $l$, which is challenging given that the calculation of EE for an interacting many-body system is already a difficult task in general. Using quantum Monte Carlo (QMC) simulations the numerical study of the scaling of 2nd R\'enyi EE has been carried out in lattice models that exhibit e.g. spontaneous O$(N)$ symmetry breaking and related critical points~\cite{Hastings2010,Humeniuk2012,Inglis2013,InglisNJP2013,KallinPRL2013,Luitz2014,KallinJS2014,Helmes2014,Laflorencie2016,JRZhao2020,JRZhao2021}, and also in models with $\Z_2$ topological order~\cite{Isakov2011,Jiang2012,Block2020}. However, despite all the progress numerical computation of EE in QMC remains difficult due to the lack of a stable estimator, especially when lattice models have multi-spin interactions or frustrations, not to mention the even more complex fermionic models where the computational complexity scales with the system size to a higher power~\cite{Grover2013,Assaad2015}. The high numerical cost of measuring EE mainly stems from the fact that to calculate the R\'enyi EE one has to enlarge the configuration space to create replicas, and perform partial connection of the partition functions between the replicas during the sampling processes.

These are the difficulties we set to overcome in this work. Here, by combining the nonequilibrium measurements of entanglement entropy based on the Jarzynski equality~\cite{Jarzynski1997,Out-of-equilibrium-2017,DEmidio2020} and the increment trick swapping replicas~\cite{Humeniuk2012,Hastings2010}, we developed a new nonequilibrium increment method, that makes use of the divide-and-conquer procedure of the nonequilibrium process to improve the speed of the simulation and the data quality of the entanglement measurement. We demonstrate the strength and versatility of the method with a few representative examples of $(2+1)d$ quantum many-body lattice model systems, in which the EEs are notoriously hard to compute. We explicitly show that previous methods are not able to extract the universal corrections beyond the leading area-law contribution in the scaling of EE, due to the lack of efficiency and precision in the EE computation.

In the QMC computation of the 2nd R\'enyi EE, the shape of spacetime configuration, as shown in Fig.~\ref{fig:fig5}, is a two-sheeted Riemann surface for 1D systems. Such a topological unit provides a stable and efficient way of carrying out the computation both conceptually and technically. In general, the n-th order R\'enyi entropy can be obtained by an n-sheeted Riemann surface, which can also help to reconstruct the low-lying entanglement spectrum for quantum many-body systems~\cite{yan2021extract}.

The structure of the paper is organized in the following manner. Firstly, we show the results of finite-size scaling of EE on three representative non-trivial systems, all in $(2+1)d$, starting from the N\'eel state of the antiferromagnetic Heisenberg model with spontaneously broken continuous symmetry, to the quantum critical point of O(3) CFT and eventually arriving at the scaling form on toric code toy model with $\Z_2$ topological order and Kagome $\Z_2$ quantum spin liquid state on a torus geometry, where in the former we also compared the results with density-matrix renormalization group (DMRG) on cylinder geometry. In both cases, the TEE of $2\gamma=2\ln(2)$ is obtained unambiguously in QMC lattice model simulations. And in all these cases, our algorithm provides reliable data with high efficiency and very large system sizes. Then we conclude the results with some immediate directions for further research. Lastly, we explain in detail the methodology of our algorithm, which is the nonequilibrium increment estimator for the R\'enyi EE~\cite{DEmidio2020,JRZhao2021} with high efficiency and precision.

\bigskip\noindent\textbf{Results}\\
In this section, we demonstrate the strength of our algorithm with three representative examples. i) Entanglement entropy at the quantum phases with spontaneous continuous symmetry breaking; ii) Entanglement entropy at the quantum critical point with conformal field theory description; iii) Entanglement entropy at the topological order phase with fractional excitations. In each case we will show that the universal correction ($s$ or $\gamma$) can be reliably determined using our method.\\

\noindent\textbf{Entanglement of spontaneous continuous symmetry breaking state.} For systems with spontaneously broken continuous symmetry, it is known that the EE follows the scaling form of Eq.~\eqref{eq:eq1} with the coefficient of the logarithmic correction $s=-N_G(d-1)/2$, where $N_G$ is the number of Goldstone modes and $d$ is the spatial dimension~\cite{Metlitski2011}.
The  logarithmic correction originates from the interplay between gapless Goldstone modes and restoration of the symmetry in a finite system.

\begin{figure}[htp!]
	\centering
	\includegraphics[width=0.98\columnwidth]{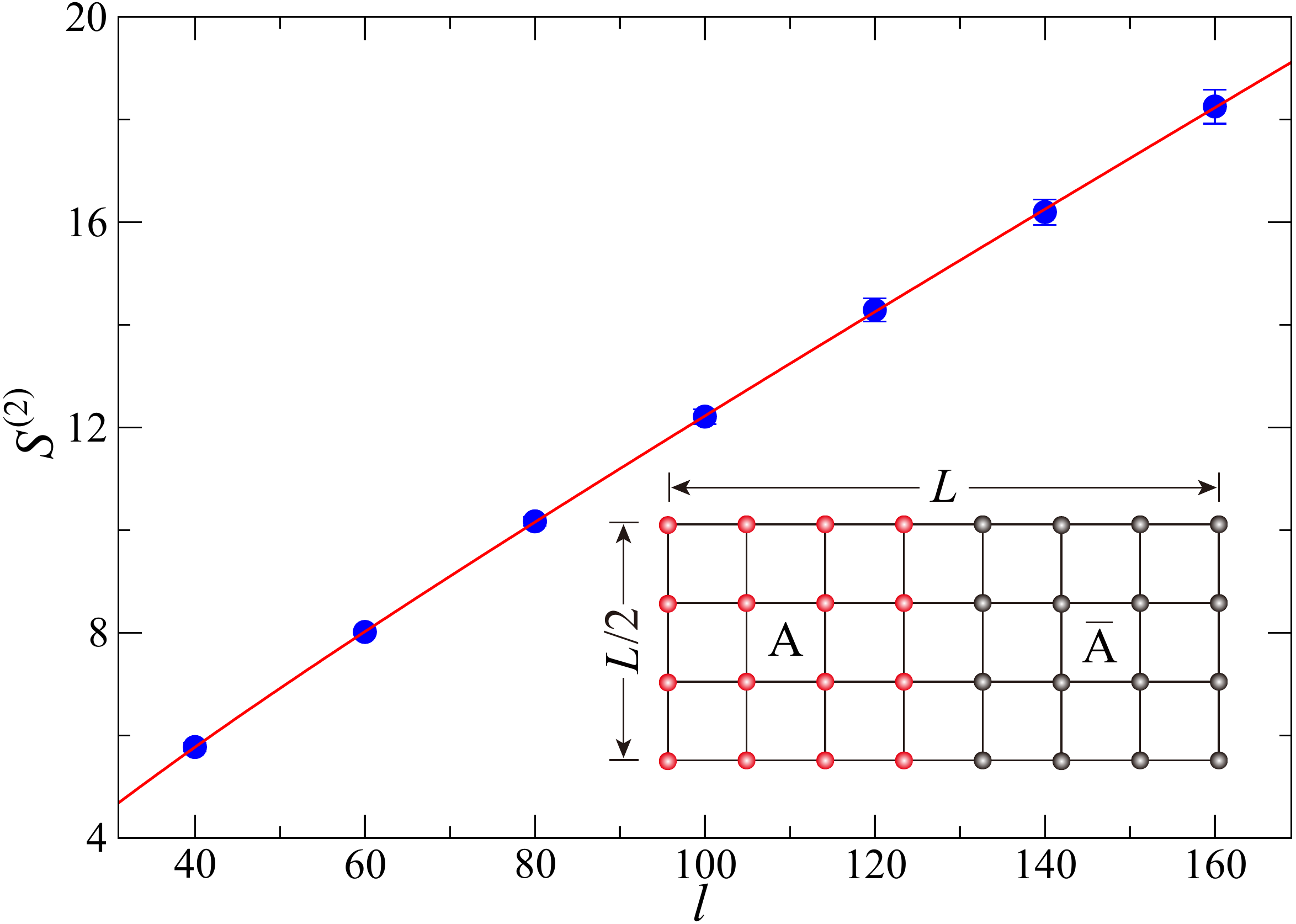}
	\caption{\textbf{Identification of Goldstone modes.} Second R\'enyi entanglement entropy of two dimensional $L\times \frac{L}{2}$ Heisenberg model for different system sizes. The inset shows the entanglement region $A$ is chosen to be a $\frac{L}{2} \times \frac{L}{2}$ cylinder on the torus without corners and with boundary length (area) $L$. The temperatures are chosen to be $1/T = L$. The fitting result is $S^{(2)}_{A}(L)=0.092(1)L+1.00(9)\ln(L)-1.63(3)$ for $L\in[40,160]$. The standard error of  the mean (SEM) is used when estimating the errors of the physical quantities.}
	\label{fig:fig1}
\end{figure}

We consider a square lattice anti-ferromagnetic Heisenberg model, whose ground state breaks the SU(2) spin rotational symmetry in the thermodynamic limit. The number of Goldstone modes is $N_G=2$ and therefore we expect $s=-1$. We measure $S^{(2)}_{A}(L)$, the 2nd R\'enyi entropy for an entangling region $A$. The simulation setup is shown in the inset of Fig.~\ref{fig:fig1}. The model is defined on a $L\times L/2$ lattice with periodic boundary conditions or a $L\times L/2$ torus, and the entangling region $A$ is chosen to be a $L/2 \times L/2$ cylinder on the torus, in such a way that there are no corners between $A$ and $\overline{A}$, and the boundary of $A$ is of length (volume) $L$. $S^{(2)}_{A} (L)$ is computed using  the finite-temperature stochastic series expansion (SSE) QMC~\cite{Sandvik1999,Syljuaasen2002} for different $L$, which is then fitted to Eq.~\eqref{eq:eq1}. From the fitting with $L\in [40, 160]$, we find $s=-1.00(9)$, in good agreement with the theoretical prediction and a previous numerical measurement with the sequential nonequilibrium method~\cite{DEmidio2020}. However, we emphasize that unlike Ref. [\onlinecite{DEmidio2020}] here we do not need to use the valence bond basis in the QMC simulation to suppress the thermal fluctuations. The more standard finite-temperature simulation $1/T=L$ in the $S^{z}$ basis already suffices to successfully achieve the desired data quality with high efficiency.\\

\begin{figure}[htp!]
	\centering
	\includegraphics[width=\columnwidth]{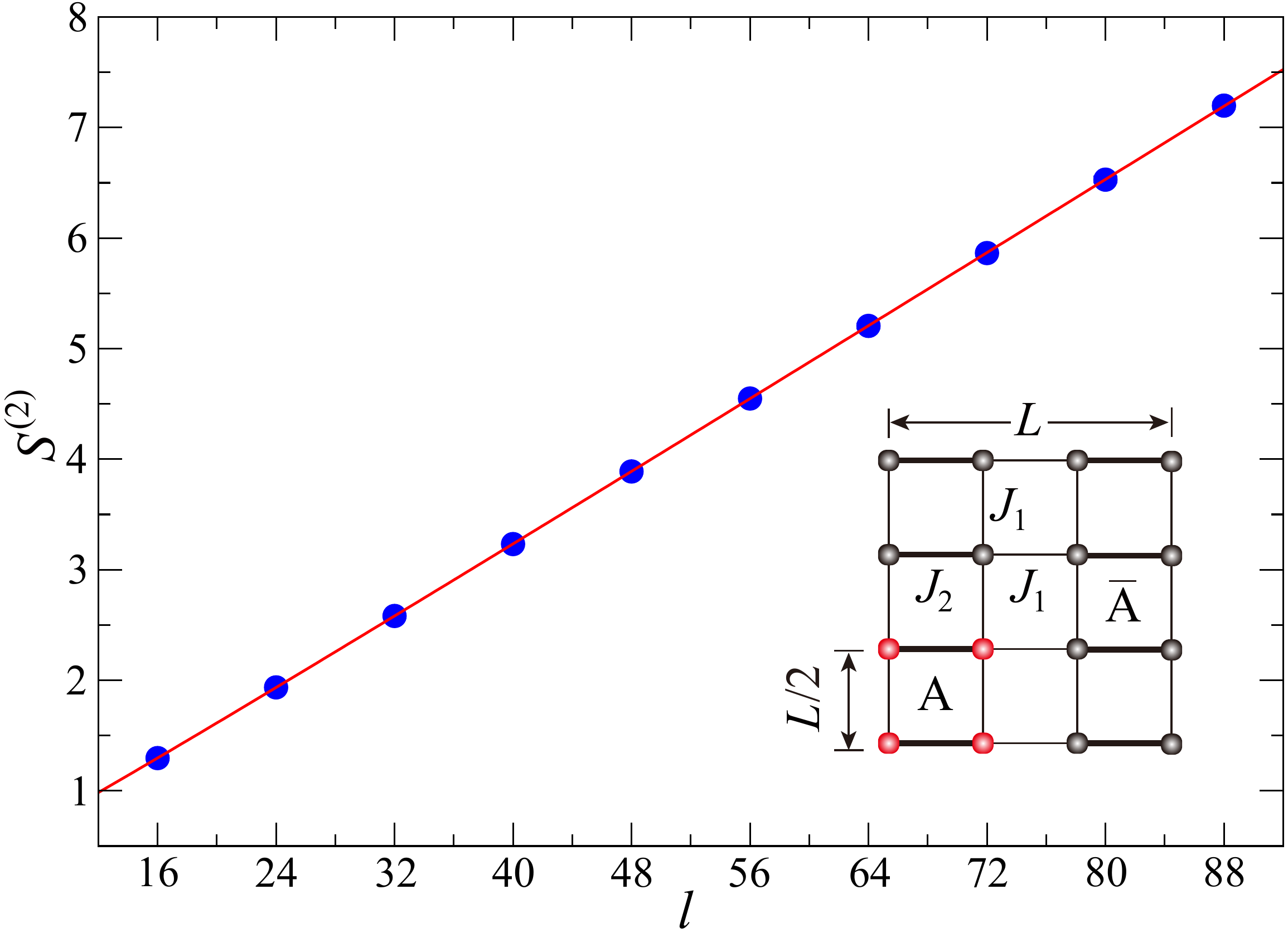}
	\caption{\textbf{Entanglement entropy at O(3) criticality.} Second R\'enyi entanglement entropy of the square lattice $J_{1}-J_{2}$ columnar dimer model for different system sizes. The entanglement region $A$ is chosen to be a $\frac{L}{2} \times \frac{L}{2}$ square with four corners and boundary length $2L$. The corners contribute to the universal log-correction with the coefficient $\beta$ related to the central charge of the underlying CFT of the O(3) transition. The fitting result is $S^{(2)}_{A}(L)=0.167(2)L-0.081(4)\ln(L)-0.124(7)$. SEM is used when estimating the errors of the physical quantities.}
	\label{fig:fig2}
\end{figure}

\noindent\textbf{Entanglement at $(2+1)d$ quantum critical point.} The second example to test the performance of our method is a lattice model that realizes the O(3) quantum critical point (QCP). For this purpose, following previous literature~\cite{NSMa2018anomalous,YCWang2021DQCdisorder,JRZhao2021}, we consider the square lattice $J_1$-$J_2$ Heisenberg model (the columnar dimer lattice model), illustrated in the inset of Fig.~\ref{fig:fig2}. The Hamiltonian reads
\begin{equation}
H_{J_1-J_2}=J_1\sum_{\langle ij \rangle} \mathbf{S}_{i}\cdot\mathbf{S}_{j}+J_2\sum_{\langle ij \rangle^{'}} \mathbf{S}_{i}\cdot\mathbf{S}_{j},
\label{eq:J1J2H}
\end{equation}
where $\langle ij \rangle$ denotes the thin $J_1$ bonds and $\langle ij \rangle^{'}$ denotes the thick $J_2$ bonds. The QCP located at $(J_2/J_1)_c=1.90951(1)$~\cite{NSMa2018anomalous} is known to fall within the $(2+1)d$ O(3) universality class. 

In the model, the presence of strong $J_2$ and weak $J_1$ bonds breaks the lattice translation symmetry. Because of the translation symmetry breaking, we choose the entangling region $A$ so that its boundary avoids strong $J_2$ bonds to reduce the finite-size error in the scaling behavior of the entanglement entropy. A similar strategy has also been adopted in recent studies of the disorder operator in the same model~\cite{YCWang2021DQCdisorder}.

We employ the torus geometry of $L\times L$ and choose a rectangle entangling region of size $L/2 \times L/2$ with four corners(shown in the inset of Fig.~\ref{fig:fig2}). We measure $S^{(2)}_{A} (L)$ at $(J_2/J_1)_c$ at $1/T=L$ and monitor its scaling behavior. 
The results are shown in Fig.~\ref{fig:fig2}. Here again, by fitting with the form in Eq.~\eqref{eq:eq1}, we obtain the coefficient of the logarithmic correction $s=0.081(4)$. This value is again consistent with the theoretical prediction for the O(3) CFT and previous numerical results on $(2+1)d$ Ising, XY and Heisenberg QCPs~\cite{Inglis2013, KallinJS2014,Helmes2014,JRZhao2020,YCWang2021DQCdisorder}. And our method can reach much larger systems sizes and has better data quality than the previous numerical results.\\

\noindent\textbf{Entanglement of topological ordered state.} In a fully gapped ground state, the entanglement entropy $S$ for a region $A$ generally satisfies $S=a l-\gamma+\cdots$ where $\gamma$ is the TEE. For simply-connected regions, $\gamma=\ln \mathcal{D}$ where $\mathcal{D}$ is the total quantum dimension of the underlying topological order. For $\Z_2$ toric code we have $\mathcal{D}=2$. Extracting TEE from numerical simulations of interacting lattice models has proven to be a challenging task.  In the following, we present our results for TEE in two different lattice models exhibiting $\Z_2$ topological order. As will be shown below, our algorithm can go beyond limitations in previous Monte-Carlo studies and the expected value of TEE is obtained in the Balents-Fisher-Girvin kagome spin-$1/2$ model, which unambiguously proves that the ground state is a $\Z_2$ spin liquid. For clarify, in this section we fix $\gamma=\ln 2$.\\

\noindent\textbf{Toy model.}

\begin{figure*}[htp!]
	\centering
	\includegraphics[width=\textwidth]{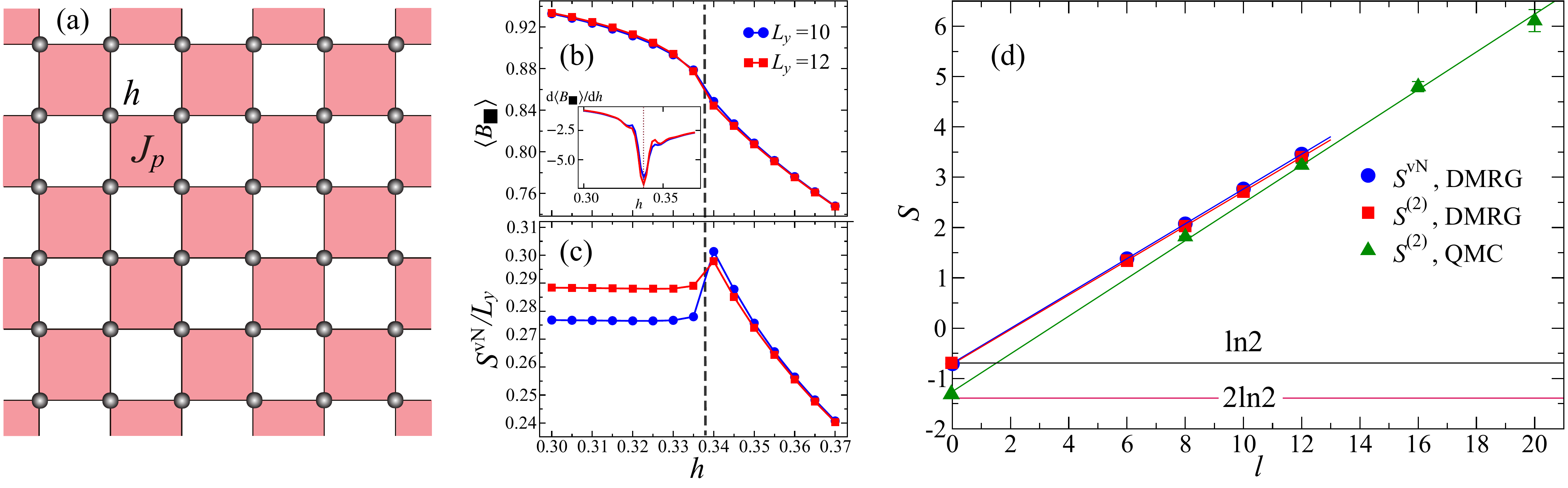}
	\caption{\textbf{Results of TEE for toy model.} (a) The checkerboard lattice $\Z_2$ gauge field model with plaquette term $B_\blacksquare$ and transverse field $h$ term as in Eq.~\eqref{eq:eq13}. (b) and (c), the DMRG results on the plaquette energy term and the von Neumann entropy for cylinder geometry with $L_x=72$ and $L_y=10$ and $12$. The deconfine-confine transition happens at $h_c=0.33$, indicating by the vertical dashed line, which is determined by the dip of the slope
		$\mathrm{d}\langle B_\blacksquare\rangle/\mathrm{d}h$ for both circumferences.
		(c) The EE at $h=0.3<h_c$ of von Neumann and 2nd R\'enyi obtained from DMRG with $72\times12$ cylinders and the QMC measurements on torus of size $L\times L$ with $L=4,6,8,10$. (d) The entangling region $A$ for QMC is $L\times L/2$ (similar with Fig.~\ref{fig:fig4} (b) and the boundary length $l=2L$.  The extrapolation for both DMRG and QMC gives rise to the TEE of values $\gamma=0.69(1)$ for cylinder geometry and $2\gamma=1.35(8)$ for torus geometry, respectively. SEM is used when estimating the errors of the physical quantities.}
	\label{fig:fig3}
\end{figure*}
To test the efficiency of our algorithm on systems with topological order, we start with a toy model of $\Z_2$ topological phase on a 2d checkerboard lattice~\cite{Assaad2016,ChuangChen2020,ChuangChen2021}, with the following Hamiltonian,
\begin{equation}
H_\mathrm{toy} = -J_p \sum_\blacksquare B_\blacksquare^{\,} - h\sum_{i} \sigma^x_i,
\label{eq:eq13}
\end{equation}
with $B_\blacksquare^{\,}\equiv\prod_{i\in\blacksquare}\sigma^z_i$ and $\blacksquare$
being the red-colored plaquettes in Fig.~\ref{fig:fig3} (a). Note that this Hamiltonian has an extensive number of conserved quantities:
\begin{equation}
	A_{\square} = \prod_{i\in \square}\sigma^x_i,
	\label{}
\end{equation}
where $\square$ are the white plaquettes in Fig.~\ref{fig:fig3}(a). In fact, $-\sum_{\blacksquare}B_{\blacksquare}-\sum_{\square}A_\square$ is the celebrated toric code model.


Below, we will combine finite-size DMRG~\cite{White1992} and QMC methods, to show  the low-energy physics of this model [c.f. Eq.\eqref{eq:eq13}] is equivalent to
the well-known toric code model under transverse field~\cite{Jiang2012},
\begin{equation*}
H_\mathrm{TCM} = -J_s \sum_\square A_\square^{\,} -J_p \sum_\blacksquare B_\blacksquare^{\,}
- h\sum_{i} \sigma^x_i,
\end{equation*}
Our DMRG simulations reveal $\langle A_\square^{\,}\rangle=1$ for the range of $h$ we consider in this section, in the
low-temperature regime of $H_\mathrm{toy}$. Thus the ground state wavefunctions of $H_\text{toy}$ and $H_{\text{TCM}}$ are \emph{identical} for $J_s>0$. In the toric code model, the transverse field induces dynamics of $\blacksquare$-plaquette excitations. As $h$ increases, such excitations condense and drive a transition to a trivial confined phase.
Thus, the toy model $H_\mathrm{toy}$ should also experience a continuous phase transition at
$h=h_c\simeq0.333$ \cite{Wu2012}, from a $\Z_2$ deconfined phase to a confined phase.

With DMRG, the ground state properties of $H_\mathrm{toy}$ are computed on a $L_x\times L_y$ cylinder geometry with  open/periodic boundary
condition along the horizontal($x$)/vertical($y$) direction.
We can directly calculate the von Neumann entropy
$$S^\mathrm{vN} = -\mathrm{tr}( \rho_\mathcal{A}\ln\rho_\mathcal{A}),$$
where the reduced density matrix
$\rho_\mathcal{A} = \mathrm{tr}_\mathcal{B} |\psi\rangle\langle\psi|$
and the subsystems $\mathcal{A}$ and $\mathcal{B}$ are both cylinders of
size $\tfrac{L_x}{2} \times L_y$.
Fig.~\ref{fig:fig3}(b) shows that the plaquette energy $\langle B_\blacksquare\rangle$-s
monotonically decrease as the transverse field $h$ grows, and a change of curve can be seen around the dashed vertical line, which highlights the point $h=h_c\simeq0.333$ (the inset shows $\frac{\mathrm{d}\langle B_\blacksquare\rangle}{\mathrm{d}h}$ and a dip is found around $h\simeq 0.338$ for both $L_y=10$ and $12$ systems). In Fig.~\ref{fig:fig3}(c), for $72\times10$ and $72\times12$ cylinders,
$S^\mathrm{vN}$-s are shown versus $h$ and exhibit peaks around $h_c$.
The above DMRG data strongly suggest the expected deconfinement-confinement transition at $h=h_c\simeq0.333$, consistent with the literature~\cite{Jiang2012,Wu2012}.

To further extract the nature of the low-$h$ phase, we perform finite-size scaling and extrapolate
the topological entanglement entropy (TEE), i.e. $\gamma$ in Eq.~\eqref{eq:eq1}, both in the cylinder geometry for DMRG and torus geometry with our QMC algorithm.
As shown in Fig.~\ref{fig:fig3} (d), we measure both von Neumann entropy $S^\mathrm{vN}$
and R\'enyi entropy $S^{(2)}$ in cylinder geometry via DMRG simulations.
With circumferences $L_y=6,8,10,12$ and fixed length $L_x=72$, the entanglement entropies
are measured and extrapolated with boundary length $l=2L_y$, where the extrapolated value of $\gamma=0.69(1)$
is obtained. For the torus geometry of size $L\times L$, the R\'enyi entropy versus
$l=2L$ is also measured in QMC simulations, and the extrapolation gives $2\gamma=1.35(8)$. Both of the results are numerically indistinguishable from $\gamma=\ln2\approx0.693$.\\

\noindent\textbf{Minimal Entropy State.}
We now discuss an important subtlety in our prescription for extracting TEE.  Since TEE is the correction to the entanglement area law, to directly extract TEE  complex prescriptions on planar partitions have been designed~\cite{Levin2006, Kitaev2006} to cancel the leading area-law contributions as well as possible corrections from corners. However, when numerically implemented on lattice such prescriptions typically suffer from severe finite-size effects because they require several non-overlapping regions.  As an example, the Levin-Wen prescription has been employed to study the TEE of the $\Z_2$ quantum spin liquid state, which in principle should yield a universal correction of $2\gamma=2\ln 2$. The best results by now only find $\ln 2$ instead of $2\ln 2$~\cite{Isakov2011,Block2020}. This is because within the allowed system sizes for large-scale numerical simulations, the size of each of the four partitions is at most $l\sim 10$, which is too small compared with the characteristic length-scale of the problem (inverse vison gap~\cite{ZY2020vison}, as discussed in the next section), and the cancellation of the entanglement entropies between different partitions are too noisy to give the TEE in a controlled manner.

To this end, the kind of entanglement cut adopted in our simulations, i.e. bipartition of a torus into two cylinders, has the advantage of maximally enlarging the entangling region while avoiding corners on the boundary. Naively, the TEE should be $2\gamma$ since the boundary has two disconnected components. However, because the boundary curve is topologically nontrivial (non-contractible cycles on a torus), the value of the TEE now depends on which ground state is used for the calculation of the entanglement entropy. More specifically, for $\Z_2$ toric code and a given entanglement cut (e.g. along the $y$ direction), only certain choices of ground states on torus yield the expected value $2\gamma=2\ln 2$. Any other ground state gives a smaller $\gamma$ and thus a larger $S$ for the same size. For this reason, these special ground states which saturate $2\gamma=2\ln 2$ are called minimum entropy states (MES)~\cite{Dong2008,ZhangYi2012}. Physically, the MES are characterized by a definite anyon flux through the non-contractible entanglement cut, so are in one-to-one correspondence with anyon excitations of the topological order.


 Numerical simulations in DMRG for the toric code toy model in the previous section and quantum spin liquid system~\cite{Jiang2012}, have shown that due to the energy minimization process in the DMRG for quasi-one-dimensional systems with finite accuracy, the algorithm actually finds the desired MES with $\gamma=\ln 2$ for the cylinder geometry (also shown in Ref.~\cite{Jiang2012}) as shown in Fig.~\ref{fig:fig7} (d).

 Our numerical results in the previous section for $\Z_2$ toric code toy model suggest that the Monte Carlo sampling processes in the our algorithm serve the same purpose in finding the expected TEE of $2\gamma=2\ln 2$, as shown in Fig.~\ref{fig:fig3}(d). In the next section, we further strengthen this point using an even more nontrivial test of the kagome $\Z_2$ spin liquid lattice model.\\

\noindent\textbf{$\Z_2$ quantum spin liquid on kagome lattice.} The last and yet the most challenging case for the entanglement entropy measurement is that of the Balents-Fisher-Girvin (BFG) model with $\Z_2$ quantum spin liquid (QSL) ground state~\cite{BFG2002,Isakov2011,Isakov2006,Isakov2012,YCWang2017QSL,GYSun2018,YCWang2018,YCWang2020,YCWang2021vestigial}. As far as we are aware of, the expected value of the TEE for such Kagome $\Z_2$ QSL model has never been observed in previous QMC computations of the 2nd R\'enyi EE.

\begin{figure*}[htp!]
	\centering
	\includegraphics[width=\textwidth]{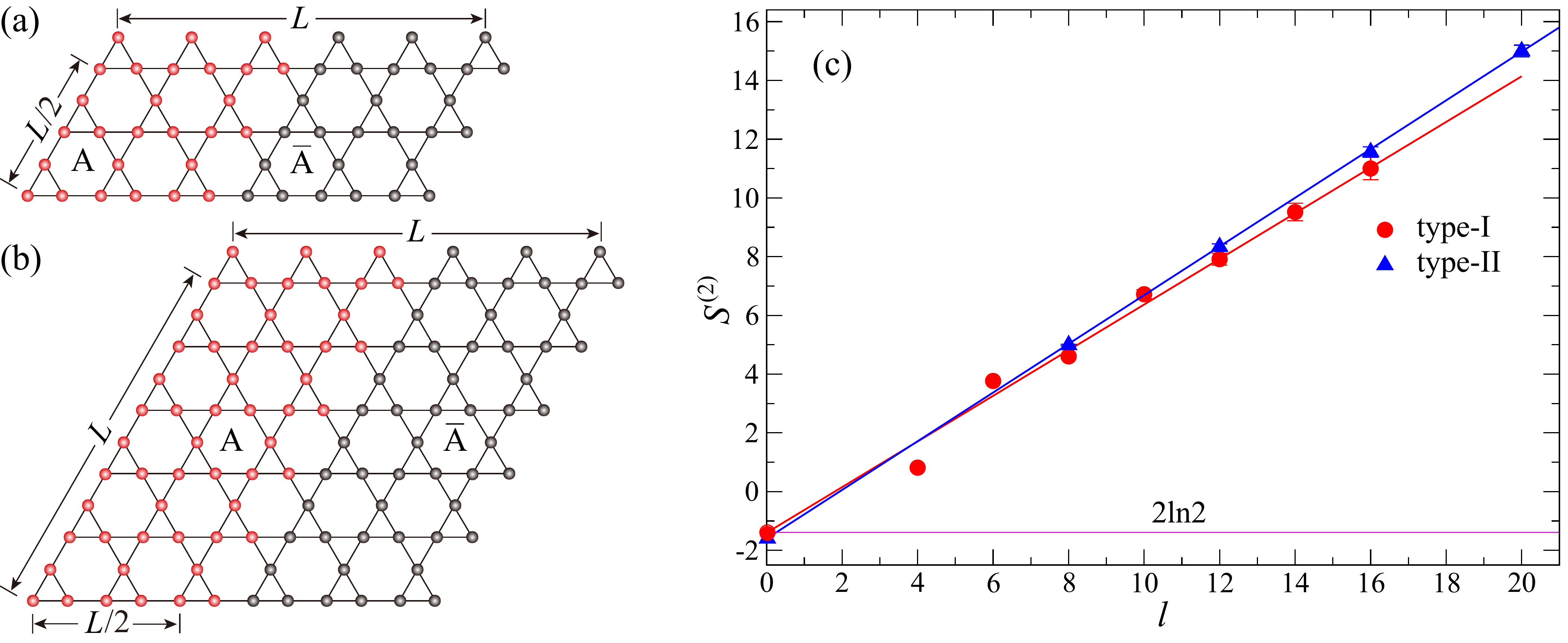}
	\caption{\textbf{Results of TEE for $\Z_2$ quantum spin liquid.} The second R\'enyi entanglement entropy of the 2D  Kagome $\Z_2$ QSL model in Eq.~\eqref{eq:eq14} for different aspect ratios, with the torus geometry of size $L\times \frac{L}{2}$ in (a) and $L\times L$ in (b), $J_{\pm}/J_z = 0.0625$ (inside the QSL phase) and $1/T=480$ (below the anyon gaps). The entangling region $A$ is chosen to be a $\frac{L}{2} \times \frac{L}{2}$ cylinder in (a) and $L\times \frac{L}{2}$ cylinder in (b), which is half of the system without corners. Therefore the entanglement entropy is expected to only have area law contribution plus a constant $\gamma$ -- the topological entanglement entropy. (c). For the $\Z_2$ QSL state on Kagome lattice with torus geometry, the $\gamma=2\ln(2)$ signifying the fractionalized statistics of the topological ordered QSL state. Our fitting results with $L=12,14,16$ in type-I geometry (a) and $L=6,8,10$ in type-II geometry (b), give rise to $\gamma=1.4(2)$. SEM is used when estimating the errors of the physical quantities.}
	\label{fig:fig4}
\end{figure*}

As illustrated in Fig.~\ref{fig:fig4} (a) and (b), the model is defined on a Kagome lattice with the following Hamiltonian
\begin{equation}
H =-J_{\pm}\sum_{\langle i,j \rangle} (S_i^{+}S_j^{-}+\text{H.c.}) + \frac{J_z}{2}\sum_{\hexagon}\Big(\sum_{i\in\hexagon} S_i^z\Big)^{2},
\label{eq:eq14}
\end{equation}
where $J_{\pm}$ is the ferromagnetic transverse nearest neighbor interaction and $J_{z}$ is the antiferromagnetic longitudinal interactions between any two spins in the hexagon of the Kagome plane.
The model is known from previous intensive QMC simulations to host a $\Z_2$ QSL and the transition from the ferromagnetic phase to the $\Z_2$ QSL occurs at $(J_{\pm}/J_z)_c=0.07076(2)$ ~\cite{Isakov2006,YCWang2020}. The identification of a $\Z_2$ QSL ground state for $(J_{\pm}/J_z) < (J_{\pm}/J_z)_c$ has been supported by several different types of measurements, such as the unconventional quantum phase transition between ferromagnetic phase and the $\Z_2$ QSL belonging to the $(2+1)d$ XY$^*$ (instead of XY) universality class~\cite{Isakov2006,Isakov2012,YCWang2018}, signifying the existence of fractional anyon excitations in the $\Z_2$ QSL phase, and the vison-pair spectra with translation symmetry fractionalization~\cite{GYSun2018}, the vestigial anyon condensation transition towards other topological ordered state~\cite{YCWang2021vestigial} and the fractional conductivity at the $(2+1)d$ XY$^*$ critical point~\cite{YCWang2020}.

This system is also relevant for on-going experimental efforts in synthesizing and identifying QSL materials in the Kagome based quantum magnets such as Zn-paratacamite  Zn$_x$Cu$_{4-x}$(OH)$_6$Cl$_2$ $(0 \le x \le 1)$ with Herbertsmithite its full Zn end~\cite{ShoresMP05,HeltonJS07,HanTH12,FuM15,NormanMR16} and Zn-doped Barlowite Zn$_x$Cu$_{4-x}$(OH)$_6$FBr and Zn-doped Claringbullite Zn$_x$Cu$_{4-x}$(OH)$_6$FCl $(0 \le x \le 1)$~\cite{FengZL17,Wen2017,ZLFeng2018PRB,ZLFeng2019,JJWen2019} with NMR~\cite{FengZL17}, neutron scattering~\cite{WeiYuan2017,YuanWei2020}, $\mu$SR~\cite{YuanWei2020} and thermodynamic measurements~\cite{WeiYuan2019,YuanWei2020nano}. Better theoretical characterization of the $\Z_2$ QSL states with fundamental probe such as the EE, and its future connection with the experimental reality such as the existences of magnetic impurities~\cite{YuanWei2020nano}, will certainly help to eventually reveal the existence of fractionalized excitations and anyonic statistics in Kagome based quantum magnets~\cite{Wen2019,Broholm2020}.

However, previous attempts of measuring the second R\'enyi EE in this model~\cite{Isakov2011} and other similar frustrated Kagome spin model~\cite{Block2020} using the Levin-Wen prescription, were not successful in revealing the actual value of the TEE $2\gamma=2\ln 2$ in Eq.~\eqref{eq:eq1}. As discussed in the previous section, although the Levin-Wen construction~\cite{Levin2006,Kitaev2006} can in principle remove the area law contribution and expose the universal constant correction of $2\gamma$, what has been observed at best is a plateau of approximately $\gamma$ for finite-size systems (e.g. $3\times8\times8$ with 3 sites per unit cell of the Kagome lattice), at an intermediate temperature below the spinon energy scale $\sim J_z$, but still comparable or higher than the vison energy scale of $\sim J^3_{\pm}/J^2_z$. It is understood that the data quality and the computational complexity in simulating even larger sizes and lower temperatures prohibit more precise determination of the 2nd R\'enyi EE of the system~\cite{Isakov2011}.

We found that our our algorithm successfully overcomes these difficulties. We consider two kinds of geometry as shown in Fig.~\ref{fig:fig4} (a) and (b), with  periodic boundary conditions in both lattice directions and aspect ratios $L_{x}/L_{y}=2$ and $L_{x}/L_{y}=1$. We choose a cylindrical entangling region $A$ without corners but with two disconnected boundaries. In this way, the second R\'enyi entropy (for a MES) should scale as
\begin{equation}
	S^{(2)}_{A}= 2\alpha L_y - 2\gamma,
\label{eq:eq14}
\end{equation}
where $2L_y$ is now the total length of the boundary and the TEE is $2\gamma=2\ln 2\approx 1.386$. We carried out the non-equilibrium increment measurement with $N=240$ parallel pieces and gradually increase the system sizes $L_x=L$ (so the two torus are $L\times L/2$ and $L\times L$, respectively), with $J_{\pm}/J_z = 0.0625$ (inside the QSL phase) and $T=1/480$ (below the anyon gaps)~\cite{GYSun2018,YCWang2020}.

The results are shown in Fig.~\ref{fig:fig4} (c). One can see that as $L$ increases, $S^{(2)}_{A}(L)$ clearly develops a linear behavior, with a converged slope ($\alpha$ in Eq.~\eqref{eq:eq14}) and more importantly, converged intercept $1.4(2)$, for both systems with different aspect ratios. It is worth noting that the converged behavior with system sizes starts to emerge at around $l \sim 12$, which may explain why previous QMC works on the same system~\cite{Isakov2011} fail to extract the value of $2\ln 2$. This observation shows the importance of having access to sufficiently large system sizes to be able to observe the full TEE. Such  large-scale computation of entanglement entropy only becomes possible because of our improved algorithm, which can be easily and robustly implemented for such complicated models, resulting in significantly improved performance. Our results thus unambiguously demonstrate the existence of $\Z_2$ QSL in the system.

As mentioned in the previous section, despite potential complications known theoretically~\cite{Dong2008,ZhangYi2012}, owing to the sampling selection mechanism built into the Monte Carlo processes, our algorithm, based on the numerical evidence, indeed finds the MES TEE, similar to the DMRG application on the other $\Z_2$ spin liquid model~\cite{Jiang2012}.

\bigskip \noindent {\bf Discussion}\\ 
The measurement of entanglement entropy differs fundamentally from more traditional probes such as order parameters, structure factors and various correlation functions, being able to reveal more subtle global information hidden in the wave function of quantum many-body systems. However the precise detection of EE has been very hard for the past decades. Here we show that previously difficult task can be greatly optimized and improved via the nonequilibrium increment method, and in this way, one can investigate the scaling behavior of EE in many $(2+1)d$ quantum many-body systems with very large system sizes, controlled errorbars and minimal computational costs. Starting from the three representative cases shown here, one can foresee the implementation and measurement of EE via  our algorithm for other topological ordered phases and phase transitions, interacting fermionic systems such as the Gross-Neveu QCPs with critical Dirac fermions~\cite{YZLiu2020}, the deconfined QED$_3$ problems of gauge fields coupled to fermion matter fields~\cite{XYXu2019,WeiWang2019,Janssen2020} and the more complicated situations of non-Fermi-liquid and quantum critical metals~\cite{Xu2017,XYXu2019Review,ZHLiu2019,XYXu2020,AKlein2020,HQYuan2020,WLJiang2021,YWu2021,YZLiu2021} and hopefully make further suggestions to the on-going experimental search for these strongly entangled quantum matter.

At last we want to mention that, except for the measurement of EE, in recent years new measurements such as the symmetry domain walls or field lines of emergent gauge field and disorder operators to probe and characterize phases and phase transitions and the associated condensation of extended objects, spontaneously breaking of the so-called higher-form symmetry~\cite{Nussinov2006, Nussinov2009, Gaiotto_2015,ji2019categorical, kong2020algebraic,JRZhao2020,XCWu2020,YCWang2021U1,Estienne2021,XCWu2021,YCWang2021DQCdisorder,TDP-Binbin-2022,low-lying-zhengyan-2021} have been successfully designed and implemented in many exotic quantum many-body systems. For example, the recent measurement of disorder operator at the deconfined quantum critical point (DQCP) has unambiguously exhibited the difference between the DQC and other QCPs with unitary CFT description~\cite{YCWang2021DQCdisorder,JRZhao2021}, as is also seen in the measurement of entanglement entropy\cite{JRZhao2021}. These new measurements are very hopeful to extract more fundamental information and lead to a deeper understanding of the interacting many body systems .

\bigskip\noindent\textbf{Methods}\\
\noindent\textbf{Replica trick.} In a quantum many-body system, the entanglement of a subsystem $A$ with the rest of the system $\overline{A}$ is most commonly identified by the von-Neumann entropy $S_{A}^{(vN)}=-\mathrm{Tr}\rho_{A}\ln\rho_{A}$. Here $\rho_{A}=\mathrm{Tr}_{\overline{A}}\rho$ is a reduced density matrix defined as the partial trace of the total density matrix $\rho$ over a complete basis of subsystem $\overline{A}$. The calculation of the von-Neumann entropy directly from the reduced density matrix in QMC simulations is difficult as usually one does not have the wave-function of a generic quantum many-body system, say, in $(2+1)d$.

\begin{figure}[htp!]
	\centering
	\includegraphics[width=0.5\columnwidth]{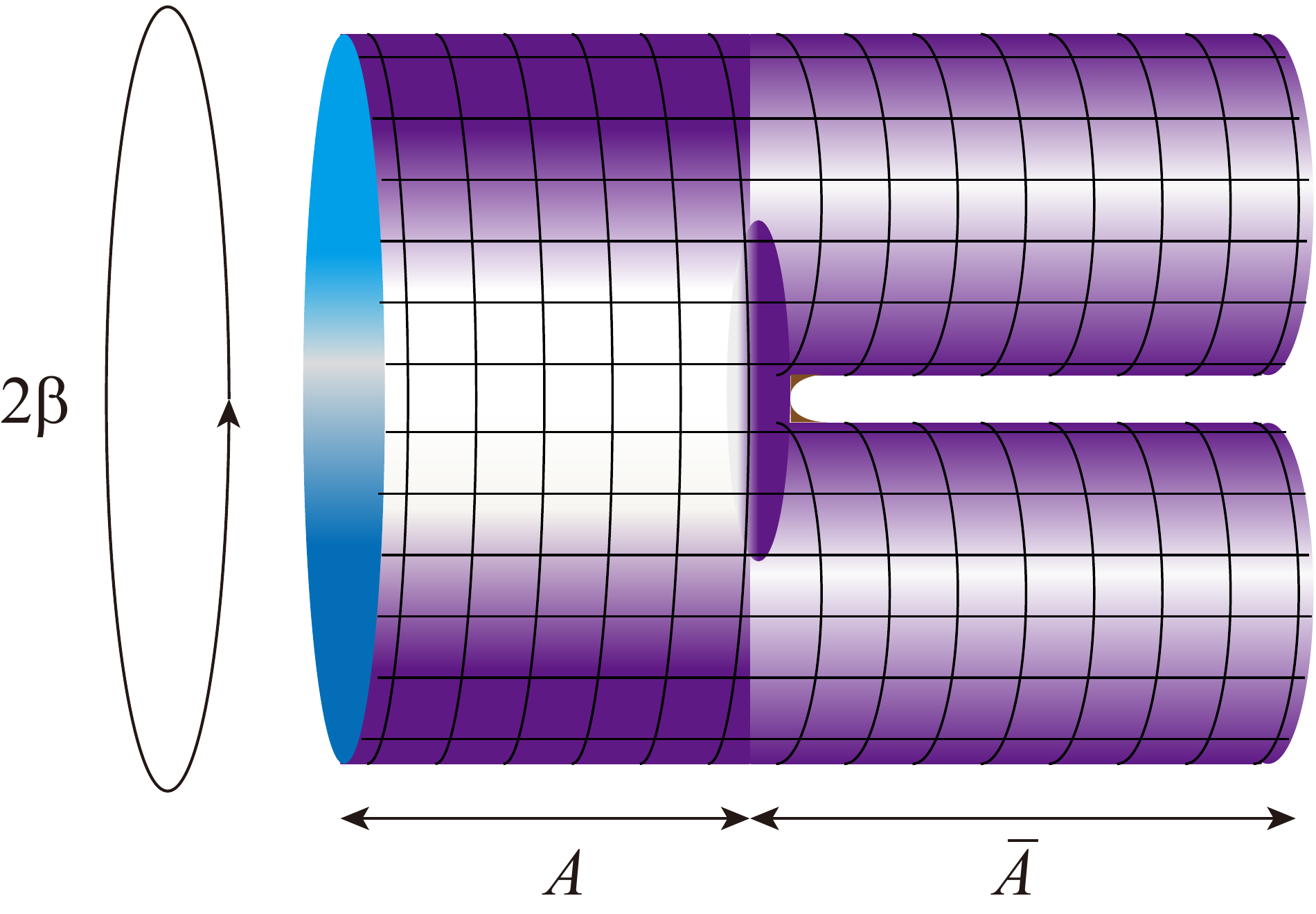}
	\caption{\textbf{The geometrical presentation of the partition function $\mathcal{Z}_{A}^{(2)}$.} Area $A$ of the two replicas is glued together, while area $\overline{A}$ of the two replicas are independent to each other. }
	\label{fig:fig5}
\end{figure}

However, the R\'enyi entropy~\cite{Calabrese_2004,Fradkin2006,Tagliacozzo2009}
\begin{equation}
\label{eq:eq2}
S^{(n)}_{A}=\frac{1}{1-n}\ln\left(\frac{\mathcal{Z}_{A}^{(n)}}{\mathcal{Z}^{(n)}}\right),
\end{equation}
which can be regarded as a generalization of the von-Neumann entropy, approaching the latter when $n\rightarrow 1$, can be calculated by the QMC method. As $\mathcal{Z}^{(n)}=[\mathrm{Tr}(e^{-\beta \mathcal{H}})]^{n}$ is the ordinary partition function of $n$ replicas of the system while $\mathcal{Z}_{A}^{(n)}$ is a modified partition function with the boundary conditions of area $A$ of the $n$ replicas changed according to the value of $n$. In the case of $n=2$ and in the 1D model, $\mathcal{Z}_{A}^{(2)}$ is two replicas with area $A$ glued together as depicted in Fig.~\ref{fig:fig5}. However for sites in $\overline{A}$ for each replica, the usual periodical boundary condition is maintained. 

Based on Eq.~\eqref{eq:eq2}, various estimators have been introduced to calculate the R\'enyi entanglement entropy\cite{Hastings2010,Humeniuk2012,Luitz2014}. However, the data quality of these estimators has severely limited their usage in higher dimensions and larger sizes. 
And for all the previous methods mentioned, the obtainable system sizes of the 2nd R\'enyi entropy of spin lattice models are very limited. There is one empirical explanation for this limitation. All the previous methods somehow measure the observable $\frac{\mathcal{Z}_{A}^{(n)}}{\mathcal{Z}^{(n)}}$ directly to obtain the R\'enyi entropy in the equilibrated ensemble. For interacting spin lattice models, $S^{(n)}_{A}$ typically obeys an area law with respect to the boundary length $l$, so $\frac{\mathcal{Z}_{A}^{(n)}}{\mathcal{Z}^{(n)}}$ should decay exponentially with the system size $L$ (typically we choose $l \propto L$ in 2D systems). As a result, the observable $\frac{\mathcal{Z}_{A}^{(n)}}{\mathcal{Z}^{(n)}}$will be too small to be obtained precisely and efficiently when we increase the system size.

 The recent proposal of the nonequilibrium method~\cite{Out-of-equilibrium-2017,DEmidio2020}, which borrows Jarzynski's equality~\cite{Jarzynski1997} of a nonequilibrium process, has made substantial progress in this regard by conducting the measurement out of the equilibrium protocol. It can obtain the 2nd R\'enyi entanglement with improved data quality for larger system sizes. The success of this method, as we will discuss in the next section, might owe to the fact that in the nonequlibrium protocol we are not taking $\frac{\mathcal{Z}_{A}^{(n)}}{\mathcal{Z}^{(n)}}$ as a direct observable in an equilibrated ensemble, but measuring the total work $W_{A}^{(n)}$ done in a nonequilibrium tunning process which is of the same order of $S_{A}^{(n)}$ according to Eq.~\ref{eq:eq12jj}.  But the disadvantage of this method, as we will show in the following sections, is the tunning process should not be too fast. However, this shortcoming can be overcome by the increment trick which gives rise to a nonequilibrium increment method, making it possible to obtain the R\'enyi entropy more precisely and efficiently  for the more complicated quantum many-body systems with frustration, multi-spin interactions, etc.We will give more details of the nonequilibrium protocol and the comparison of our method with the original proposal in the subsequent sections.\\

\noindent\textbf{Nonequilibrium measurement.} In this section the  nonequilibrium method~\cite{DEmidio2020} will be reviewed as the foundation of our further development. The first step of the nonequilibrium method is to introduce a partition function $\mathcal{Z}_{A}^{(n)}(\lambda)$ parameterized by $\lambda$.
$\mathcal{Z}_{A}^{(n)}(\lambda)$ is defined as the sum of a collection of partition functions $Z_{B}^{(n)}$ weighted by a binomial distribution $g_{A}\left(\lambda, N_{B}\right)=\lambda^{N_{B}}(1-\lambda)^{N_{A}-N_{B}}$. According to this definition, $\mathcal{Z}_{A}^{(n)}(\lambda)$ can be written as
\begin{equation}
\begin{split}
\mathcal{Z}_{A}^{(n)}(\lambda)&=\sum_{B \subseteq A} \lambda^{N_{B}}(1-\lambda)^{N_{A}-N_{B}} Z_{B}^{(n)}\\
&=\sum_{B \subseteq A} g_{A}\left(\lambda, N_{B}\right) Z_{B}^{(n)}
\end{split}
\end{equation}
where $\mathcal{Z}_{A}^{(n)}(1)=\mathcal{Z}_{A}^{(n)}$ and $\mathcal{Z}_{A}^{(n)}(0)={\mathcal{Z}_{\varnothing}^{(n)}}$. With this definition, the R\'enyi entanglement entropy can be rewritten as $S_{A}^{(n)}=-\frac{1}{n-1}\frac{\mathcal{Z}_{A}^{(n)}(1)}{\mathcal{Z}_{A}^{(n)}(0)}$, which leads to an integral expression
\begin{equation}
S_{A}^{(n)}=\frac{1}{1-n} \int_{0}^{1} d \lambda \frac{\partial \ln \mathcal{Z}_{A}^{(n)}(\lambda)}{\partial \lambda}
\label{eq:eq4}
\end{equation}
where $\frac{\partial \ln \mathcal{Z}_{A}^{(n)}(\lambda)}{\partial \lambda}$ can be measured by $\langle \frac{\partial \ln {g}_{A}}{\partial \lambda} \rangle_\lambda=\langle (N_B/\lambda)-(N_A-N_B)/(1-\lambda) \rangle_\lambda$. Ref.~\cite{DEmidio2020} puts forward a nonequilibrium process where the system evolves from a configuration of $\mathcal{Z}_{\varnothing}^{(n)}(\mathcal{Z}_{A}^{(n)}(0))$ to a configuration of $\mathcal{Z}_{A}^{(n)}(\mathcal{Z}_{A}^{(n)}(1))$ through  tunnning from $\lambda=0$ to $\lambda=1$. The total work done in this process is
\begin{equation}
W_{A}^{(n)}=-\frac{1}{\beta} \int_{t_{i}}^{t_{f}} d t \frac{d \lambda}{d t} \frac{\partial \ln g_{A}\left(\lambda(t), N_{B}(t)\right)}{\partial \lambda} 
\label{eq:eq5}
\end{equation}
where $\lambda(t_{i})=0$ and $\lambda(t_{f})=1$.

According to the Jarzynski's equality~\cite{Jarzynski1997}, the free energy difference $\Delta F$ accumulated in such a path can be extracted by the ensemble average of nonequilibrium measurement by the following relation:
\begin{equation}
\Delta F=-\beta^{-1}\ln\overline{e^{-\beta W}},
\end{equation}
the overbar refers to the average of an ensemble of nonequilibrium paths. As the free energy for a canonical ensemble can be expressed by $F=-\beta^{-1}\ln \mathcal{Z}$, the R\'enyi entanglement entropy can then be estimated by
\begin{equation}
S_{A}^{(n)}=\frac{1}{1-n} \ln \left(\left\langle e^{-\beta W_{A}^{(n)}}\right\rangle\right)
\label{eq:eq12jj}
\end{equation}
where $S_{A}^{(n)}$ is rewritten in the free energy difference of ensemble $\mathcal{Z}_{A}^{(n)}(0)$ with ensemble $\mathcal{Z}_{A}^{(n)}(1)$.

Note that although $\lambda(t)$ can take different forms, in this paper we only consider the case where $\lambda$ varies uniformly with $t$.  In this  case, $W_{A}^{(n)}$ can be calculated by accumulating $\Delta \ln g_{A}(\lambda(t_{m}),N_{B}(t_{m}))= \ln g_{A}(\lambda(t_{m+1}),N_{B}(t_{m}))-\ln g_{A}(\lambda(t_{m}),N_{B}(t_{m}))$ or alternatively recording the value of $\frac{g_{A}(\lambda(t_{m+1}),N_{B}(t_{m}))}{g_{A}(\lambda(t_{m}),N_{B}(t_{m}))} $ at each time and  multiplying them at the end of the measurement.\\

\noindent\textbf{Nonequilibrium increment method.} Above is the basic concept of the nonequilibrium measurement, our optimization of this method takes advantage of the fact that the partition function $ \mathcal{Z}_{A}^{(n)}(\lambda) $ is well defined at every $\lambda \in [0,1]$, so that we can insert a set of identity elements $\frac{\mathcal{Z}_{A}^{(n)}(kh)}{\mathcal{Z}_{A}^{(n)}(kh)} (k=1,2,...,N-1)$ and rewrite $\frac{\mathcal{Z}_{A}^{(n)}(1)}{\mathcal{Z}_{A}^{(n)}(0)}$ as
\begin{equation}
\frac{\mathcal{Z}_{A}^{(n)}(1)}{\mathcal{Z}_{A}^{(n)}(0)}=\frac{\mathcal{Z}_{A}^{(n)}(Nh)}{\mathcal{Z}_{A}^{(n)}((N-1)h)}\frac{\mathcal{Z}_{A}^{(n)}((N-1)h)}{\mathcal{Z}_{A}^{(n)}((N-2)h)}\cdots\frac{\mathcal{Z}_{A}^{(n)}(h)}{\mathcal{Z}_{A}^{(n)}(0)}
\label{eq:eq8}
\end{equation}
where $N$ is an integer and $h=\frac{1}{N}$.  The increment trick is discussed in Ref.~\cite{Humeniuk2012} as a way to overcome the limitation of equilibrium measurement of the R\'enyi entanglement entropy with large entangling regions. Our algorithm is the nonequilibrium version of the increment trick. According to Eq.~\eqref{eq:eq8} and Eq.~\eqref{eq:eq4} the R\'enyi entanglement entropy can be reexpressed as
\begin{equation}
S_{A}^{(n)}=\frac{1}{1-n}\sum_{k=0,1,\cdots,N-1} \int_{kh}^{(k+1)h} d \lambda \frac{\partial \ln \mathcal{Z}_{A}^{(n)}(\lambda)}{\partial \lambda}.
\label{eq:eq9}
\end{equation}

\begin{figure}[htp!]
	\centering
	\includegraphics[width=\columnwidth]{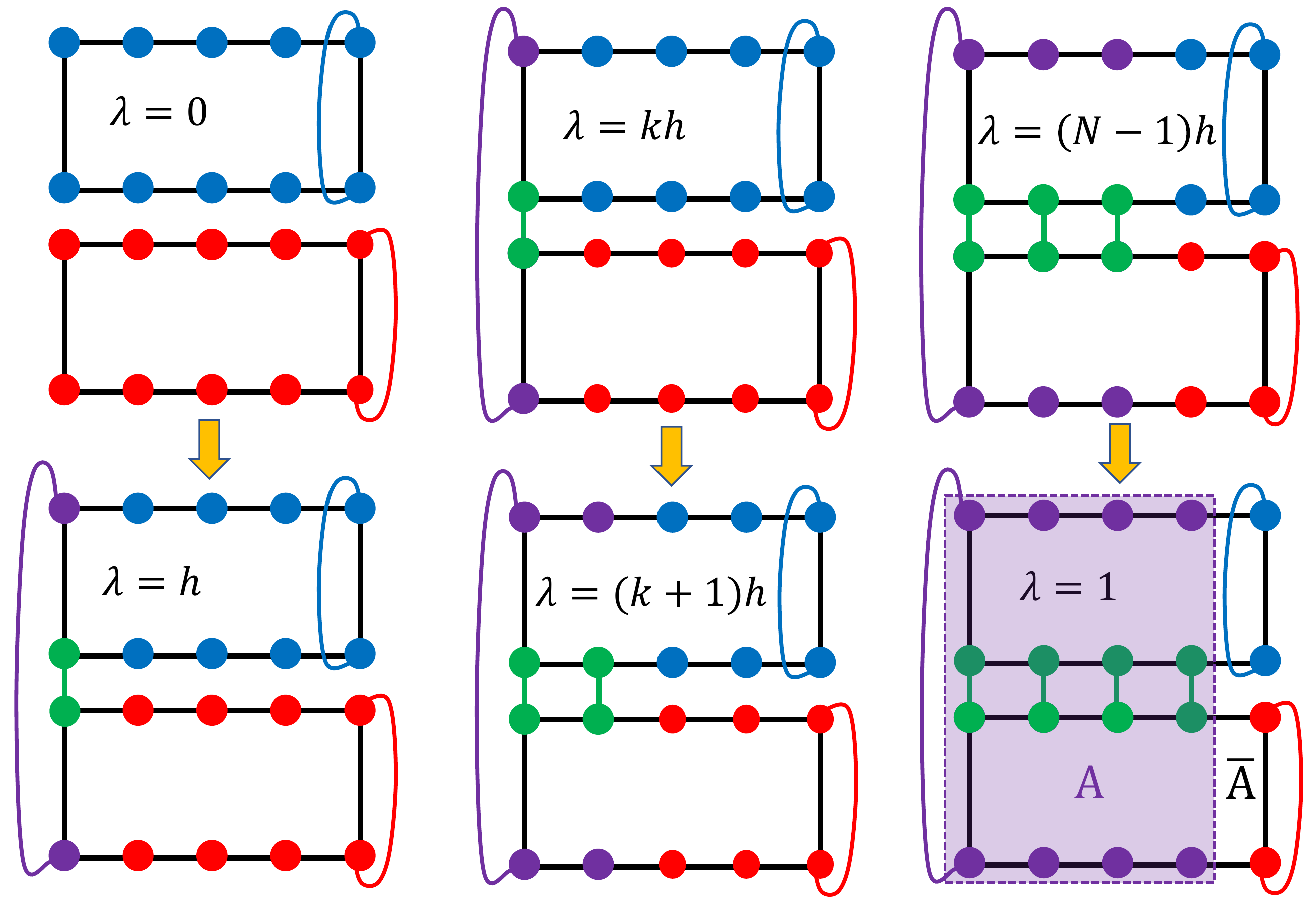}
	\caption{\textbf{A schematic plot of the nonequilibrium increment algorithm}. It splits a consecutive $\lambda-$parameterized nonequilibrium process into $N$ independent smaller pieces. For each piece we start from a thermalized state of the partition function $\mathcal{Z}_{A}^{(2)}(\lambda=kh)$ and carry out the nonequilibrium measurement from $\lambda=kh$ to $\lambda=(k+1)h$. For the $k=0$ piece (the left column), the starting configuration is two independent replicas with ordinary periodical conditions and as the system evolves to $\lambda=h$ the configuration becomes two modified replicas with some sites in region $A$ are glued together. The $k=h$ (the middle column) and $k=(N-1)h$ (the right column) pieces are carried in parallel. The final entanglement entropy is obtained from the summation of these independent pieces, as shown in Eq.~\eqref{eq:eq10}.}
	\label{fig:fig6}
\end{figure}

Jarzynski's equality~\cite{Jarzynski1997} can be carried out on each piece $\frac{\mathcal{Z}_{A}^{(n)}((k+1)h)}{\mathcal{Z}_{A}^{(n)}(kh)} (k=1,2,...,N-1)$ and the corresponding small piece of integral $\int_{kh}^{(k+1)h} d \lambda \frac{\partial \ln \mathcal{Z}_{A}^{(n)}(\lambda)}{\partial \lambda}$. Then we obtain
\begin{equation}
S_{A}^{(n)}=\frac{1}{1-n}\sum_{k=0,1,\cdots,N-1} \ln \left(\left\langle e^{-\beta W_{k,A}^{(n)}}\right\rangle\right)
\label{eq:eq10}
\end{equation}
where $W_{k,A}^{(n)}$-s are defined same as Eq.~\eqref{eq:eq5} but $\lambda(t_{i})=kh$ and $\lambda(t_{f})=(k+1)h$ for each small piece in the nonequilibrium process, as shown in Fig.~\ref{fig:fig6}. In this way, our algorithm follows the protocol:
\begin{enumerate}
	\item We perform quantum Monte Carlo simulation on the partition function $\mathcal{Z}$ and store the thermalized QMC configuration (one replica) for later use.
	\item Prepare two replica configurations as the thermalized configuration of $\mathcal{Z}^{(2)}_{\varnothing}$ and then send them to
	$N$ parallel processes.
	\item As shown in Fig.~\ref{fig:fig6}, for process $k$ we set the initial value of $\lambda$ to be $\lambda(t_{i})=kh$. The value of $\lambda$ controls the probability of sites in $A$ (the entanglement region) joining or leaving the glued geometry of the replicas, with the following probabilities:
	\begin{equation}
	\label{eq:eq11}
	P_{\text{join}}=\min\{\frac{\lambda}{1-\lambda},1\} \quad P_{\text{leave}}=\min\{\frac{1-\lambda}{\lambda},1\}.
	\end{equation}
	
	Each MC sweep then consists of the following steps:
	\begin{itemize}
		\item Each site in $A$ can choose whether to stay or leave the region according to Eq.~\eqref{eq:eq11}. After the decision is made, change the topology, i.e. update the connectivity of the entangling region $A$. 
		\item After the trace structure is determined, carry out the MC updates on the replicas.
	\end{itemize}
	\item At this step, we fix $\lambda=kh$ for process $k$ and conduct several MC sweeps to thermalize the configuration at the beginning of the nonequilibrium measurement.
	\item Start the nonequilibrium measurement. Increase the value of $\lambda$ by $\Delta \lambda$ and record the value of $ \frac{g_{A}(\lambda(t_{m+1}),N_{B}(t_{m}))}{g_{A}(\lambda(t_{m}),N_{B}(t_{m}))} $ . Here $\lambda(t_{m})=kh+m\Delta \lambda$ and $\lambda(t_{0})=\lambda(t_{i}=kh)$. Then carry out a MC sweep. Repeat this process until $\lambda(t_{m})$ reaches the value of $(k+1)h$. For each process, $\left \langle e^{-\beta W_{k,A}^{(n)}}\right\rangle = \left \langle \prod_{m=0}^{h/\Delta \lambda-1} \frac{g_{A}(\lambda(t_{m+1}),N_{B}(t_{m}))}{g_{A}(\lambda(t_{m}),N_{B}(t_{m}))} \right\rangle$.
	\item In the end, we collect the observable from all the $N$ parallel processes and sum them to get the R\'enyi entanglement entropy according to Eq.~\eqref{eq:eq10}.
\end{enumerate}

\begin{figure}[htp!]
	\centering
	\includegraphics[width=\columnwidth]{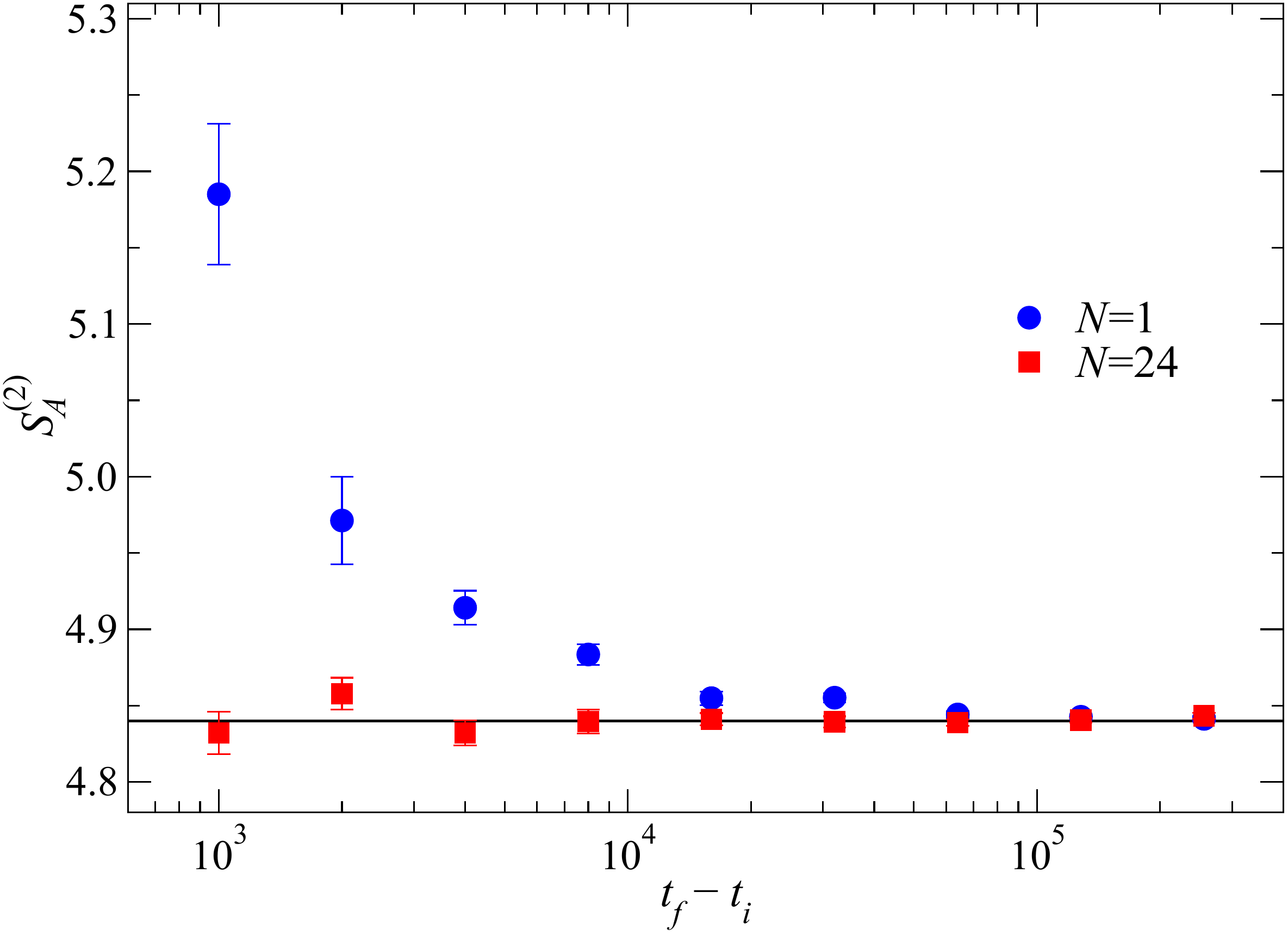}
	\caption{\textbf{Convergence of EE for different quench time.} The 2nd R\'enyi entanglement entropy $S^{(2)}_{A}$ of a 2D antiferromagnetic Heisenberg model on a $32\times 16$ torus versus the quench time $t_{f}-t_{i}$ for the nonequilibrium increment method with the total number of nonequilibrium pieces  taken to be $N=1$ and $N=24$. The entangling region $A$ is a $16\times 16$ lattice which is chosen as depicted in the inset of Fig.~\ref{fig:fig1}. SEM is used when estimating the errors of the physical quantities.}
	\label{fig:fig7}
\end{figure}

To test the efficiency of our algorithm with the original nonequilibrium proposal, we carry out the measurement of the second R\'enyi EE of a 2D antiferromagnetic Heisenberg model on a $L_{x}:L_{y}=32\times16$ torus, with the partition of $A$ and $\overline{A}$ chosen to be the same as the inset of Fig.~\ref{fig:fig1}. Note that when the total number $N$ of small processes equals 1, our algorithm reduces to the original proposal of the last section. Fig.~\ref{fig:fig7} shows the comparison of the results of $S^{(2)}_{A}$ versus the total quench time for $N=1$ and $N=24$. For the $N=1$ case, one sees that as $t_{f}-t_{i}$ becomes larger, the second R\'enyi EE gradually converges. When $t_{f}-t_{i}$ is not sufficiently large a shift of the R\'enyi EE from the converged value of $S^{(2)}_{A}=4.84(1)$ manifests. We find that this deviation is systematic and is probably caused by two reasons. First, when the quench is not slow enough, not all sites in $A$ will join the topology of glued geometry at the end of the nonequilibrium process, so the final state of the tunning process is not a configuration of $\mathcal{Z}_{A}^{2}(1)$ which gives rise to a bias in Eq.~\ref{eq:eq12jj}. Second, the observable of the nonequilibrium protocol is actually an integral defined in Eq.~\ref{eq:eq5}, and in the algorithm the sample mean method is used to estimate this integral, which naturally brings a systematic error that decays when the quench time increases. Although the errors mentioned above can always be eliminated by increasing the quench time, in practice it is not an ideal way because of the limited computing time. This defies the controlled computation of R\'enyi EE of larger and more complicated systems such as the Kagome spin model.

 For the $N=24$ case, the 2nd R\'enyi entropy converges faster and has smaller errorbars than the results of $N=1$ as the quench time $t_{f}-t_{i}$ increases. This is because the total quench time from $\lambda=0$ to $\lambda=1$ is actually $24\times(t_{f}-t_{i})$ in the simulation. The parallelization of the nonequilibrium process increases the total quench time by $N$ times, which actually cures the problem arising from not enough quench time.  One thing we want to stress is, for the comparison in Fig.~\ref{fig:fig7}, the total number of the bins for $N=1$ case is 24 times bigger than that of $N=24$ case, this ensures that the total computing power and simulation time for the two cases are nearly the same. The comparison shows that our method cures the limitation of the original algorithm and has smaller errorbars and better convergence.

We would like to end this section by mentioning that the nonquilibrium increment measurement of entanglement, implemented here and in Ref.~\cite{DEmidio2020}, could also be applied on other physics systems beyond quantum many-body lattice model and even condensed matter physics~\cite{Shirts2003,Palassini2011}.

\section*{Acknowledgement}
J.R.Z., B.B.C., Z.Y. and Z.Y.M. would like to thank Jonathan D'Emidio for encouraging and fruitful discussion on the details of the quench time of the Jarzynski estimator, and its applications in other subjects. They also thank Chenjie Wang for valuable discussions on the theoretical meaning of MES and related issues. They acknowledge support from the RGC of Hong Kong SAR of China (Grant Nos.  17303019, 17301420, 17301721 and AoE/P-701/20), the Strategic
Priority Research Program of the Chinese Academy of
Sciences (Grant No. XDB33000000), the K. C. Wong
Education Foundation (Grant No. GJTD-2020-01) and
the Seed Funding "Quantum-Inspired explainable-AI" at
the HKU-TCL Joint Research Centre for Artificial Intelligence.
Y.C.W. acknowledges the supports from the NSFC under Grant Nos.~11804383 and 11975024, and the Fundamental Research Funds for the Central Universities under Grant No. 2018QNA39. M.C. acknowledges support from NSF under award number DMR-1846109. We thank the Computational Initiative at the Faculty of Science and the Information Technology Services at the University of Hong Kong and the Tianhe platforms at the National Supercomputer Centers in Tianjin and Guangzhou for their technical support and generous allocation of CPU time.
The authors also acknowledge Beijng PARATERA Tech CO.,Ltd.(\url{https://www.paratera.com/}) for providing HPC resources that have contributed to the research results reported within this paper.


\end{document}